\documentclass[conference]{IEEEtran}
\usepackage[shortlabels]{enumitem}
\usepackage{graphicx} 
\usepackage{multicol}
\usepackage{epsfig} 
\usepackage{amsmath,amsthm} 
\usepackage{amssymb}  
\usepackage{float}
\usepackage{cases,setspace,adjustbox}
\usepackage{cite}
\usepackage{algorithmic}
\usepackage[ruled,vlined]{algorithm2e}
\usepackage{array}
\usepackage[update,prepend]{epstopdf}
\usepackage{eqparbox}
\usepackage{fixltx2e}
\usepackage{url}
\usepackage{amsthm}
\usepackage{array}
\usepackage{caption}
\usepackage{xcolor}
\usepackage{mathtools}
\usepackage{bbm}
\usepackage{multirow}
\newcommand{\vect}[1]{\boldsymbol{#1}}
\usepackage{booktabs}

\usepackage[font=normal]{subfig}
\newcommand{\pidle}{p_{\text{I}}}
\newcommand{\psucci}[1]{p_{\text{S}}^{(#1)}}
\newcommand{\pbusy}[1]{p_{\text{S}}^{(#1)}}
\newcommand{\pcol}{p_{\text{C}}}
\newcommand{\psucc}{p_{\text{S}}}

\newcommand{\lidle}{\sigma_\text{I}}
\newcommand{\lsucc}{\sigma_\text{S}}
\newcommand{\lcol}{\sigma_\text{C}}

\newcommand{\AoI}[1]{\Delta_{#1}}		
\newcommand{\initAoIT}[2]{\delta^{-}_{#1{#2}}} 
\newcommand{\AoIT}[2]{{\Delta}_{#1{#2}}}
\newcommand{\AvgAoIT}[2]{\widetilde{\Delta}_{#1{#2}}}
\newcommand{\AvginitAoIT}[2]{{\mathbf{\Delta}}^{-}_{#1{#2}}}

\newcommand{\NA}{\mathrm{N}}

\newcommand{\T}{\mathcal{T}}
\newcommand{\I}{\mathcal{I}}

\theoremstyle{definition}

\newtheorem{proposition}{Proposition}

\newtheorem{corollary}{Corollary}

\newcommand{\SG}[1]{{\color{black} {#1}}}

\begin{document}
\title{A Non-Cooperative Multiple Access Game for Timely Updates} 
\author{Sneihil Gopal$^{*}$, Sanjit K. Kaul$^{*}$, Rakesh Chaturvedi$^{\dagger}$, Sumit Roy$^{\ddagger}$\\
$^{*}$Wireless Systems Lab, IIIT-Delhi, India\\
$^{\dagger}$Department of Social Sciences \& Humanities, IIIT-Delhi, India\\
$^{\ddagger}$University of Washington, Seattle, WA\\
\{sneihilg, skkaul, rakesh\}@iiitd.ac.in, sroy@uw.edu}
\maketitle
\begin{abstract}
We consider a network of selfish nodes that would like to minimize the age of their updates at the other nodes. The nodes send their updates over a shared spectrum using a CSMA/CA based access mechanism. We model the resulting competition as a non-cooperative one-shot \SG{multiple access} game and investigate equilibrium strategies for two distinct medium access settings (a) collisions are shorter than successful transmissions and (b) collisions are longer. We investigate competition in a CSMA/CA slot, where a node may choose to transmit or stay idle. We find that medium access settings exert strong incentive effects on the nodes. We show that when collisions are shorter, transmit is a weakly dominant strategy. This leads to all nodes transmitting in the CSMA/CA slot, therefore guaranteeing a collision. In contrast, when collisions are longer, no weakly dominant strategy exists and under certain conditions on the ages at the beginning of the slot, we derive the mixed strategy Nash equilibrium. 
\end{abstract}
\section{Introduction}
The Internet-of-Things (IoT) will have large numbers of devices sense and communicate information (either their own status or that of their proximate environment) to a network coordinator/aggregator or to other devices. Applications include real-time monitoring for disaster management, environmental monitoring, industrial control and surveillance
, which require timely delivery of updates to a central station. Another set of popular applications include vehicular networking for future autonomous operations where each vehicular node broadcasts a vector (e.g. position, velocity and other status information) to enable applications like collision avoidance and platooning. 

Given the many applications, it is essential to investigate spectrum sharing by nodes that would like freshness of information. We measure freshness using the age of information~\cite{kaul2012real} metric. We consider $N$ selfish nodes that share spectrum via a CSMA/CA (carrier sensed multiple access with collision avoidance) based access mechanism. Each node would like to minimize the age of its status at the other nodes in the network.


We model the competition for the shared spectrum as a non-cooperative one-shot \SG{multiple access} game \SG{parameterized by the age of every node and the medium access settings. This is motivated by the following consideration. The interaction for spectrum access is most realistically modeled as a repeated game~\cite{zuhan} in which the ages of the nodes constitute the state of the system that endogenously evolve as a result of players' strategies over time~\cite{SGAoI2019}. Methodologically, however, it is important to understand the one-shot game and its equilibria first before delving into a detailed analysis of the repeated game. 
Therefore, in this work, we restrict ourselves to a one-shot game that is played by the nodes in a CSMA/CA slot. 

We assume that each node knows the ages of status updates at the beginning of the slot and can choose either to transmit during a slot or stay idle. We investigate the equilibrium strategies for two distinct medium access settings. Specifically, we consider when a collision (which is when multiple transmissions overlap resulting in all being decoded in error) is shorter than the length of a successful (interference free) data transmission. This is, for example, the case when networks use the RTS/CTS based access mechanism defined for the $802.11$ MAC~\cite{bianchi}. Alternately, a collision may be at least as long as the length of a successful transmission. For example, when networks use the basic access mechanism~\cite{bianchi} of the $802.11$ MAC.}

\SG{We find that medium access settings exert strong incentive effects on the nodes.} We show that when the collision slot is smaller than the successful transmission slot, transmitting during a slot is a weakly dominant strategy. This \SG{result} 
is independent of the initial ages. This, of course, leads to wastage of the shared spectrum and the age of updates of none of the nodes is reduced at the end of the slot. On the other hand, \SG{the access setting where} collision is longer, no dominant strategy exists. 
\SG{In this case}, we analytically derive the mixed strategy Nash equilibrium for when the ages at the beginning of the slot satisfy certain conditions. 

\SG{Our work provides insights into how competing nodes that value timeliness would share the spectrum under different medium access settings. Specifically, our results indicate that under decentralized decision making by nodes, the access setting with longer successful transmissions is more vulnerable to collisions than the other.} 

The rest of the paper is organized as follows. Section~\ref{sec:related} discusses the related works followed by the network model in Section~\ref{sec:model}. The game is described in Section~\ref{sec:game} and the equilibrium strategies are derived in Section~\ref{sec:eq_strategy}. Results are discussed in Section~\ref{sec:results}. We conclude in Section~\ref{sec:conclusions}.
\section{Related Work}
\label{sec:related}

Unlike age, throughput as the payoff function in wireless networks has been extensively studied from the game theoretic point of view (see~\cite{jin2002,mario2005,chen2010}). 
Age of information has been investigated for networks with multiple users sharing a slotted system 
in~\cite{Kaul-Yates-isit2017,kosta2019age,Hsu-isit2018,farazi2019fundamental,maatouk2020optimality,chen2019age}. In~\cite{Kaul-Yates-isit2017} and~\cite{kosta2019age} authors considered scheduled and random access mechanisms. In~\cite{Hsu-isit2018} authors studied a scheduling problem with respect to age and in~\cite{farazi2019fundamental} authors studied a multi-source multi-hop wireless network with age as a performance metric. In~\cite{maatouk2020optimality} authors investigated age in a CSMA-based network and formulated an optimization problem to minimize the total average age of the network. In~\cite{chen2019age} authors considered the problem of minimizing the age over a random access channel and proposed distributed age-efficient transmission policies. 

In~\cite{impact2017,YinAoI2018,SGAoI2018,SGAoI2019} authors studied games with age as the payoff function. In~\cite{impact2017} and~\cite{YinAoI2018}, authors studied an adversarial setting where one player aims to maintain the freshness of information updates while the other player aims to prevent this. In earlier work~\cite{SGAoI2018}, we proposed a game theoretic approach to study the coexistence of Dedicated Short Range Communication (DSRC) and WiFi, where the DSRC network desires to minimize the average age of information and the WiFi network aims to maximize the average throughput. 
In~\cite{SGAoI2019}, via the repeated game model we were able to shed better light on the interaction of age and throughput optimizing networks. 
Unlike these works, here we provide insights into how competing nodes that value timeliness of their information at others would share the spectrum using a CSMA/CA based medium access.

\section{Network Model}
\label{sec:model}
Our network consists of $N$ nodes, indexed $1,2,\ldots,N$, which contend for access to a shared wireless medium. Each node uses a CSMA/CA based access mechanism. We assume that all nodes can sense each other's packet transmissions. This allows modeling the CSMA/CA mechanism as a slotted multiaccess system. A slot may either be (a) an idle slot in which no node transmits a packet, (b) a successful transmission slot in which exactly one node transmits an update packet that is decoded successfully by all other nodes, (c) a collision slot in which more than one node transmits a packet and as a result none of the packets are successfully decoded. We assume that all nodes always have a fresh status update packet to send. Each node would like to minimize the age of its status update packets at the other nodes in the network.

Let $\tau_i$ denote the probability with which node $i$ attempts transmission in a slot. Let $\pidle$ be the probability of an idle slot. We have
\begin{align}
\pidle = \prod\limits_{i=1}^{\NA} (1-\tau_i).
\label{Eq:probidle}
\end{align}
Let $\psucci{i}$ be the probability of a successful transmission by node $i$ in a slot and let $\psucc$ be the probability of a successful transmission in a slot. We say that node $i$ sees a busy slot if in the slot node $i$ doesn't transmit and exactly one other node transmits. Let $\pbusy{-i}$ be the probability that a busy slot is seen by node $i$. Let $\pcol$ be the probability that a collision occurs in a slot. We have
\begin{align}
&\psucci{i}= \tau_i\prod\limits_{\substack{j=1\\j\neq i}}^{\NA} (1-\tau_j),\enspace \psucc= \sum\limits_{i=1}^{\NA}\psucci{i},\nonumber\\
&\psucci{-i}= \sum\limits_{\substack{j=1\\j\neq i}}^{\NA}\tau_j\prod\limits_{\substack{k=1\\k\neq j,k\neq i}}^{\NA}(1-\tau_k)\text{\ and\ }\pcol = 1-\pidle-\psucc.
\label{Eq:probsucc}
\end{align}

Let $\sigma_I, \sigma_S$ and $\sigma_C$ denote the lengths of an idle, successful, and collision slot, respectively. Next we define the age of a node's status updates at other nodes in terms of the above probabilities and slot lengths.
\subsection{Age of a Node's Information}
\label{sec:age}
\begin{figure}[t]
	\begin{center}
		\includegraphics[width=\columnwidth]{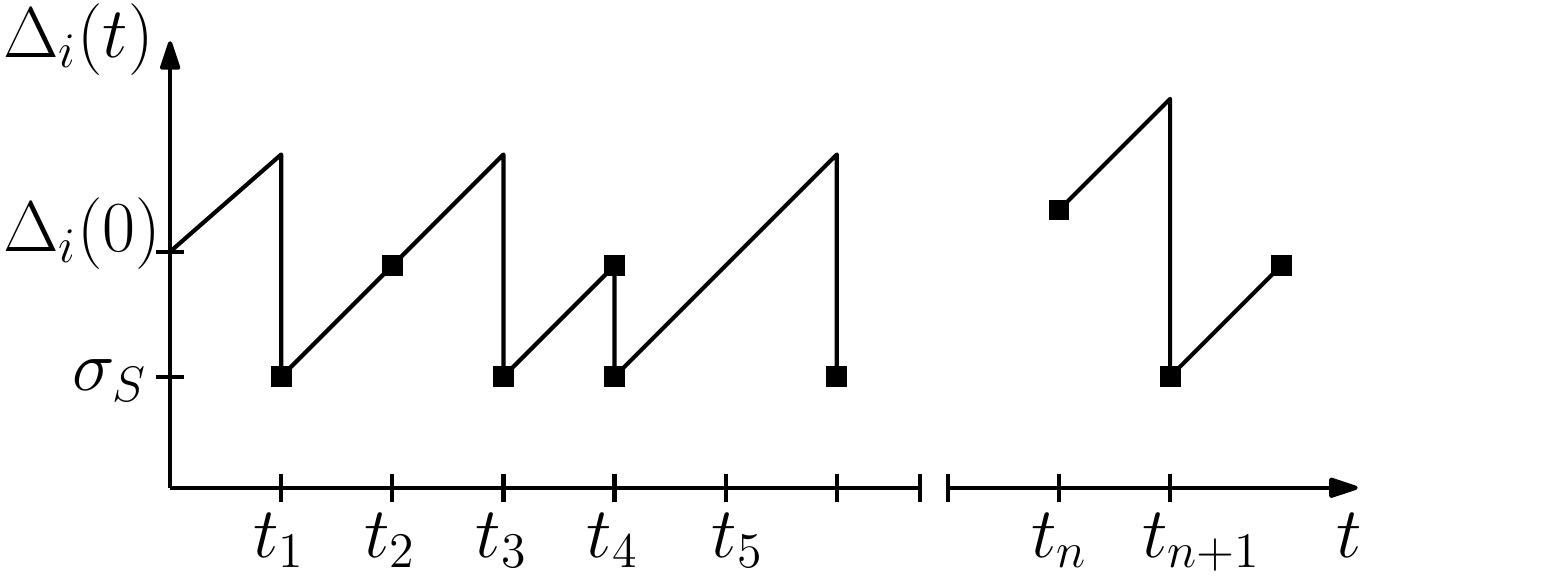}
	\end{center}
	\caption{\small Sample path of age $\AoI{i}(t)$ of node $i$'s updates at other nodes in the network. $\AoI{i}(0)$ is the initial age. A successful transmission resets the age to $\lsucc$. The time instants $t_{n}$, where, $n \in \{1,2,\dots\}$, show the slot boundaries. The slot lengths are determined by the type of slot.}
	\label{fig:instaAoI}%
\end{figure}

\begin{figure*}[t]
\centering
\includegraphics[width = 0.45\textwidth]{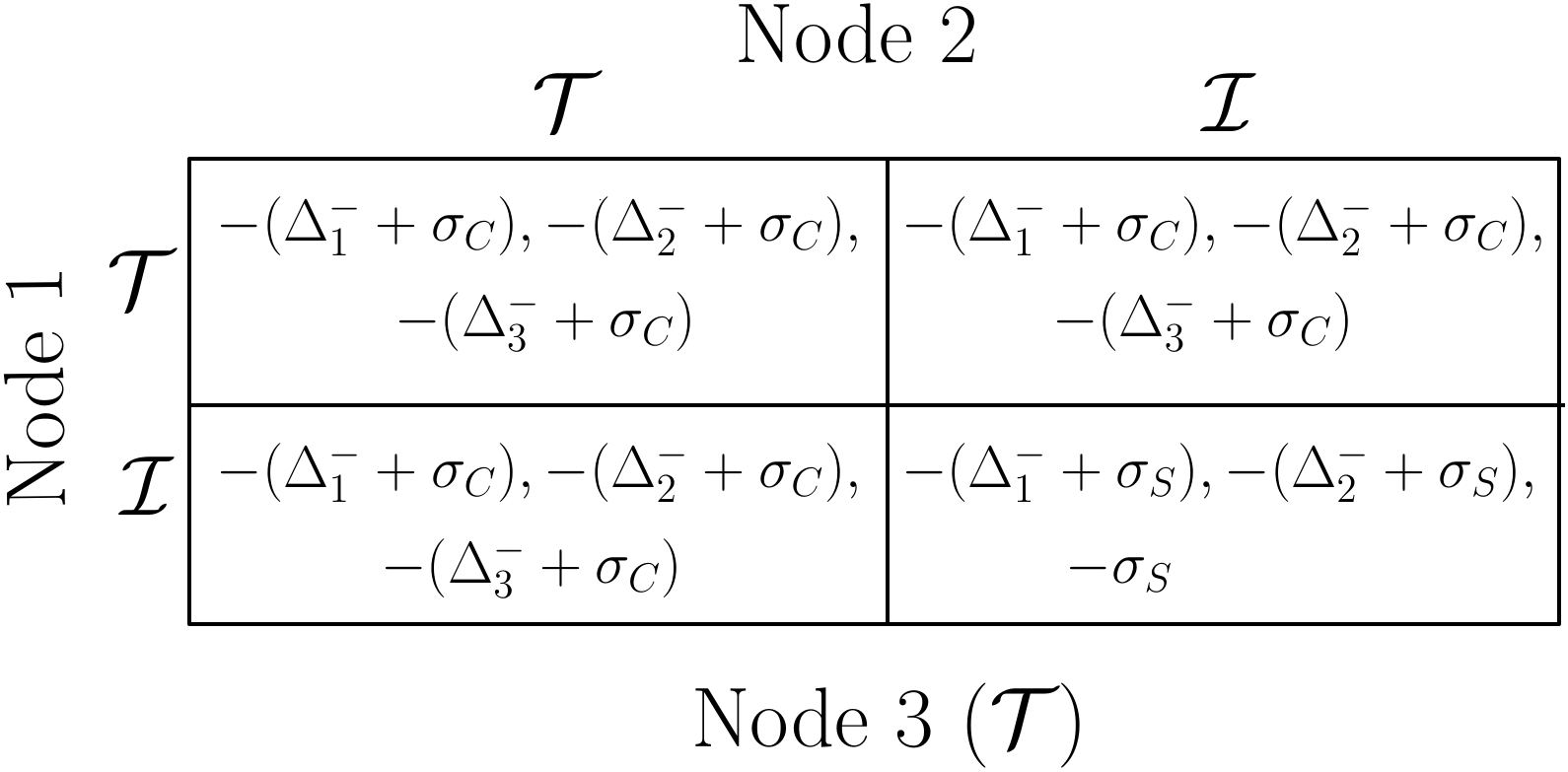}
\enspace
\includegraphics[width = 0.45\textwidth]{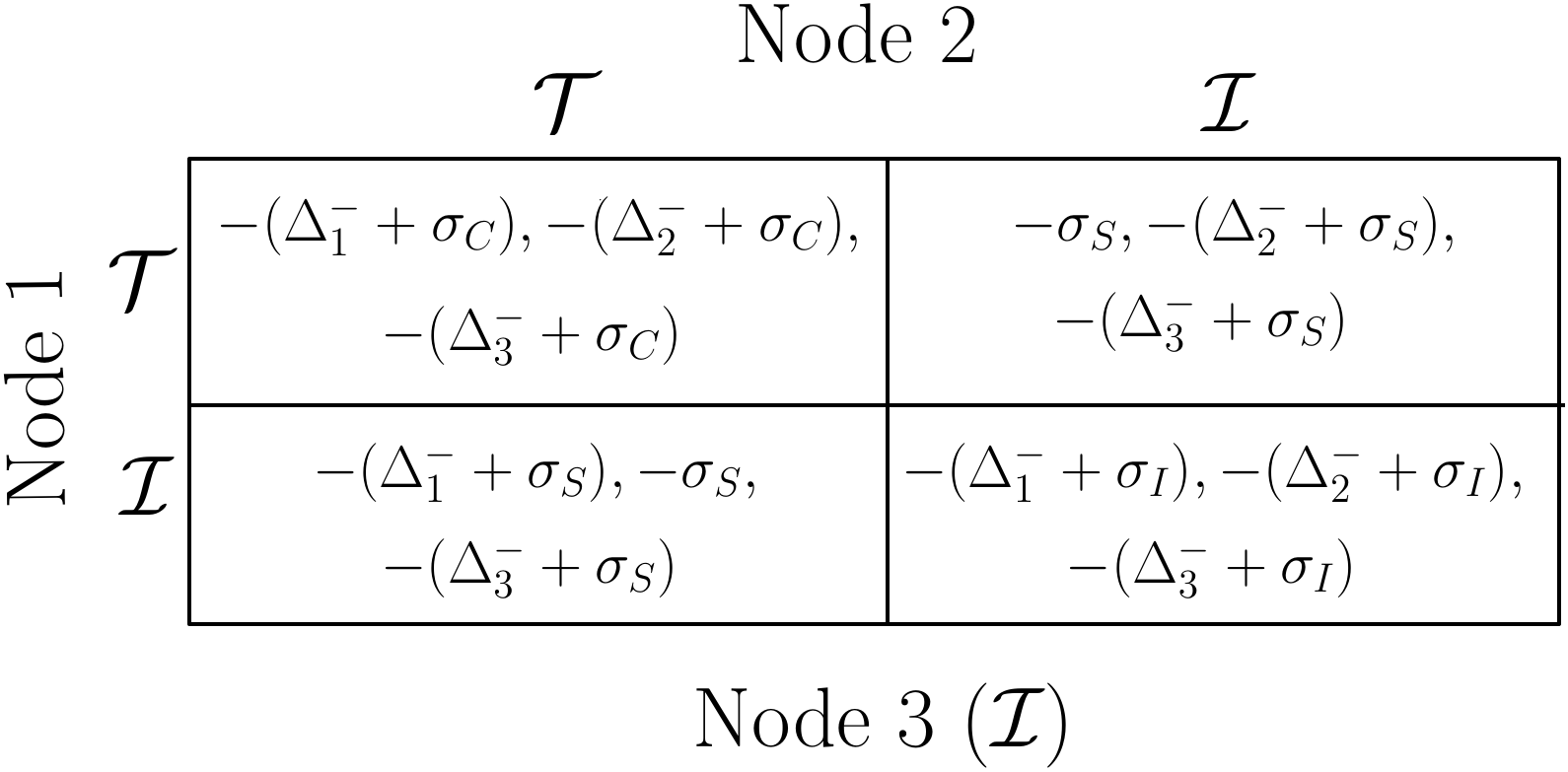}
\caption{Payoff matrix with slot lengths and age values for the game $G$ when $3$ nodes contend for the medium. Nodes can choose between transmit ($\mathcal{T}$) and idle ($\mathcal{I}$).}
\label{fig:payoff_3player}
\end{figure*}
Let $u_{i}(t)$ be the timestamp of the most recent status update of a node $i$ at other nodes $j\ne i$ at time $t$. The age of node $i$'s update at any other node $j$ at time $t$ is $\Delta_{i}(t) = t - u_{i}(t)$~\cite{kaul2012real}. We assume that a status update packet that node $i$ attempts to transmit in a slot contains an update of age $0$ at the beginning of the slot. 

Note that node $i$'s age at the end of a slot is determined by its age at the beginning of the slot and the type of the slot. As a result, node $i$'s age at any other node $j$ either resets to $\lsucc$ if a successful transmission occurs or increases by $\lidle$, $\lcol$ or $\lsucc$ at all other nodes, respectively, when an idle slot, collision slot or a busy slot occurs. Figure~\ref{fig:instaAoI} shows an example sample path of the age $\AoI{i}(t)$. In what follows we will drop the explicit mention of time $t$ and let $\Delta^{-}_{i}$ and $\AoIT{i}{}$, respectively, be the age of node $i$'s update at the beginning and end of the slot.

The age $\AoIT{i}{}$ at the end of a slot is thus a random variable whose conditional PMF given $\Delta^{-}_{i}$ is
\begin{align}
P[\AoIT{i}{} = \delta_{i}|\Delta^{-}_{i} = \initAoIT{i}{}] = 
	\begin{cases}
	\pidle & \delta_{i} = \initAoIT{i}{}+\lidle ,\\
	\pcol & \delta_{i} = \initAoIT{i}{}+\lcol,\\
	\psucci{-i} & \delta_{i} = \initAoIT{i}{}+\lsucc,\\
	\psucci{i} & \delta_{i} = \lsucc,\\
	0 & \text{otherwise.}
	\end{cases}
	\label{Eq:AoIPMF}
\end{align}
Using~(\ref{Eq:AoIPMF}), we define the conditional expected age
\begin{align}
\AvgAoIT{i}{} &\overset{\Delta}{=} E[\AoIT{i}{} = \delta_{i}|\Delta^{-}_{i} = \initAoIT{i}{}].\nonumber\\
&=(1-\psucci{i})\initAoIT{i}{}+(\pidle\lidle+\psucc\lsucc+\pcol\lcol).
\label{Eq:AoI}
\end{align}
\section{Game Model}
\label{sec:game}
We define a one-shot \SG{multiple access} game to model the interaction between the nodes in a slot. In every CSMA/CA slot, nodes must compete for access with the goal of minimizing their age. We capture the interaction in a slot as a non-cooperative parameterized strategic one-shot \SG{multiple access} game $G = (\mathcal{N},(\mathcal{S}_k)_{k\in\mathcal{N}},(u_k)_{k\in\mathcal{N}},\AvginitAoIT{}{})$, where $\mathcal{N}$ is the set of players, $\mathcal{S}_k$ is the set of strategies of player $k$, $u_k$ is the payoff of player $k$ and $\AvginitAoIT{}{}$ is the additional parameter input to the game $G$, which is the vector of ages of the nodes' updates ($\Delta^{-}_{i}$ for node $i$) at the beginning of the slot. We define the game $G$ next. 
\begin{itemize}
\item \textbf{Players:} The set of players $\mathcal{N} = \{1,2,\dots,\NA\}$ is simply the set of all nodes in the network.
\item \textbf{\SG{Strategy Space}:} Let $\T$ denote transmit and $\I$ denote idle. The set of pure strategies for node $i$ is $\mathcal{S}_{i} = \{\T,\I\}$. 
We allow nodes to play mixed strategies. Define $\boldsymbol{\Phi}_{i}$ as the set of all probability distributions over the set of strategies $\mathcal{S}_{i}$ of player $i$. A mixed strategy for player $i$ is an element $\phi_{i} \in \boldsymbol{\Phi}_{i}$, where $\phi_{i}$ is a probability distribution over $\mathcal{S}_{i}$. We require $\phi_{{i}}(s_{{i}})\geq 0$ for all $s_{{i}} \in \mathcal{S}_{{i}}$ and $\sum_{s_{{i}} \in \mathcal{S}_{{i}}} \phi_{{i}}(s_{{i}}) = 1$. 
\SG{In this work, since the strategy spaces are binary, a mixed strategy for node $i$ is identified by specifying a probability $\tau_i \in [0,1]$ with which $i$ attempts transmission in a slot. As a result, the probability distribution corresponding to node $i$ is $\phi_{{i}} = \{\phi_{{i}}(\T), \phi_{{i}}(\I)\} = \{\tau_{i}, 1-\tau_{i}\}$.}

\item \textbf{Payoffs:} We can calculate the probabilities~(\ref{Eq:probidle})-(\ref{Eq:probsucc}) for each node in the network. The probabilities when substituted in~(\ref{Eq:AoIPMF}) can be used to calculate the average age in~(\ref{Eq:AoI}). 
\SG{For every node $i$, its expected payoff when its own transmission probability is $\tau_i$ and the vector of others' transmission probabilities is $\vect{\tau}_{-i}$ is given by}
\begin{align}
u_{i}(\tau_{i},\vect{\tau}_{-i}) = -\AvgAoIT{i}{}(\tau_{i}, \vect{\tau}_{-i}). 
\label{Eq:agepayoff}
\end{align}
Each node would like to maximize its payoff, which is the same as minimizing its expected age at the end of the stage.
\end{itemize}
\section{Equilibrium Strategies}
\label{sec:eq_strategy}
We will separately consider two common 
\SG{medium access} settings found in CSMA/CA (a) $\lcol<\lsucc$ and (b) $\lcol \ge \lsucc$. The first setting is akin to RTS/CTS (request-to-send and clear-to-send) based medium access defined in the $802.11$ MAC. The use of short RTS/CTS messages to reserve the medium before accessing it to send a larger data payload (status update) packet reduces the average length of a collision slot. The second setting is akin to the basic access mechanism defined in the $802.11$ MAC and doesn't use RTS/CTS. All collisions are between packets carrying data payloads.

\subsection{When $\lcol \le \lsucc$}
Proposition~\ref{prop:lsuccgrlcol} summarizes the strategy of choice for nodes in the network when the collision slots are shorter than or equal to successful transmission slots. 
\begin{proposition}
When $\lcol\le\lsucc$, for the one-shot \SG{multiple access} game $G$, transmit ($\T$) is a weakly dominant strategy. 

\begin{proof}[\textbf{Proof:}]
As stated in~\cite{zuhan}, $s_{i}\in S_{i}$ is a weakly dominant strategy for player $i$, if for any possible combination of the other players' strategies, player $i$'s payoff from $s_{i}$ is weakly more than that from $s_{i}'$. That is, $u_{i}(s_{i},s_{-i})\geq u_{i}(s'_{i},s_{-i}), \forall s_{-i}\in S_{-i}$.

\SG{We verify that for every node $i$, pure strategy $\T$ weakly dominates the pure strategy $\I$. We fix a node $i$. Any vector of pure strategies $\vect{s}_{-i}$ played by nodes other than node $i$ can be assigned to one of the following cases: ($1$) exactly one of the other nodes chooses to transmit while others stay idle or, ($2$) more than one of the other nodes chooses to transmit, or ($3$) all other nodes choose to stay idle. We now consider these cases in detail and illustrate each of them using the payoff matrix of a $3$-player game shown in Figure~\ref{fig:payoff_3player}.}

\textit{Case $1$: Exactly one of the other nodes chooses to transmit ($\T$):} In this case, if node $i$ chooses to transmit, a collision slot occurs and its age becomes $\Delta_{i} = \Delta_{i}^{-}+\lcol$. On the other hand, if node $i$ chooses to stay idle, it will see a busy slot and age at the end of the slot will be set to $\Delta_{i} = \Delta_{i}^{-}+\lsucc$. In this case, if $\lcol<\lsucc$, node $i$ will choose to transmit ($\T$) and if $\lcol=\lsucc$ node $i$ will be indifferent between idle and transmit. The above argument can be demonstrated using the payoff matrix of a 3-player game (see Figure~\ref{fig:payoff_3player}). Assume node $2$ chooses transmit (idle) and node $3$ chooses idle (transmit). In this case, transmit is a weakly dominant strategy for node $1$ if $u_{1}(\T,\T,\I)\geq u_{1}(\I,\T,\I)$ and $u_{1}(\T,\I,\T)\geq u_{1}(\I,\I,\T)$. As shown in Figure~\ref{fig:payoff_3player}, $u_{1}(\T,\T,\I) = u_{1}(\T,\I,\T) = -(\Delta_{1}^{-}+\lcol)$ and $u_{1}(\I,\T,\I) = u_{1}(\I,\I,\T) = -(\Delta_{1}^{-}+\lsucc)$. When $\lcol<\lsucc$, $u_{1}(\T,\T,\I)> u_{1}(\I,\T,\I)$ and $u_{1}(\T,\I,\T)> u_{1}(\I,\I,\T)$, whereas, when $\lcol = \lsucc$, $u_{1}(\T,\T,\I)= u_{1}(\I,\T,\I)$ and $u_{1}(\T,\I,\T)= u_{1}(\I,\I,\T)$. 

\textit{Case $2$: More than one of the other nodes choose to transmit ($\T$):} 
In this case, collision will occur. Irrespective of the choice made by node $i$, $\Delta_{i} = \Delta_{i}^{-}+\lcol$. Player $i$ is indifferent between idle ($\I$) and transmit ($\T$) when $\lcol\le\lsucc$. This corresponds to the case in the 3-player game when node $2$ and node $3$ choose transmit and node $1$ will prefer transmit ($\T$) if $u_{1}(\T,\T,\T)\geq u_{1}(\I,\T,\T)$. As shown in Figure~\ref{fig:payoff_3player}, $u_{1}(\T,\T,\T) = -(\Delta_{1}^{-}+\lcol)$ and $u_{1}(\I,\T,\T) = -(\Delta_{1}^{-}+\lcol)$. Clearly, $u_{1}(\T,\T,\T)= u_{1}(\I,\T,\T)$ and node $1$ is indifferent between transmit and idle. 

\textit{Case $3$: All other nodes choose to stay idle ($\I$):} 
In this case, suppose node $i$ along with other nodes chooses to stay idle too. We have $\Delta_{i} = \Delta_{i}^{-}+\lidle$. In case node $i$ transmits, its status update will be successfully received and its age at all other nodes will reset to $\Delta_{i}=\lsucc$. Further note that the age of a node's status at other nodes at the beginning of the slot is at least as large as the length $\lsucc$ of a successful transmission slot. This is because $\lsucc$ is the time a fresh update must age before it is successfully received by another node. Thus $\Delta_{i}^{-}+\lidle > \lsucc$ and the node will prefer to transmit. In the $3$-player game, for this case, node $1$ will prefer transmit if $u_{1}(\T,\I,\I)\geq u_{1}(\I,\I,\I)$. We have $u_{1}(\T,\I,\I) = - \lsucc$ and $u_{1}(\I,\I,\I) = -(\Delta_{1}^{-}+\lidle)$. Thus, node $1$ chooses transmit. 

\end{proof}
\label{prop:lsuccgrlcol}
\end{proposition}

\subsection{When $\lcol > \lsucc$}
\SG{For this medium access setting, no weakly dominant strategy exists and we look for mixed strategies.}
\begin{proposition}
For the one-shot \SG{multiple access} game $G$, when $\lcol>\lsucc$ no \SG{weakly} dominant strategy exists.

\begin{proof}[\textbf{Proof:}]
Similar to Proposition~\ref{prop:lsuccgrlcol}, we assume that in a slot, for different selections of strategies by other nodes in the network, node $i$ may choose either to transmit ($\T$) or stay idle ($\I$), depending on which strategy gives a higher payoff. To find the strategy that is beneficial for node $i$, we consider the following 
cases: ($1$) exactly one of the other nodes chooses to transmit while others stay idle, or ($2$) all other nodes choose to stay idle, or ($3$) more than one of the other nodes chooses to transmit.  

\textit{Case $1$: Exactly one of the other nodes chooses to transmit ($\T$):} 
In this case, if node $i$ chooses to transmit, a collision slot occurs and its age becomes $\Delta_{i} = \Delta_{i}^{-}+\lcol$. In contrast, if node $i$ chooses to stay idle, it will see a busy slot and age at the end of the slot will be set to $\Delta_{i} = \Delta_{i}^{-}+\lsucc$. Since, $\lcol>\lsucc$, node $i$ will choose to stay idle ($\I$).


\textit{Case $2$: All other nodes choose to stay idle:} 
In this case if node $i$ chooses to stay idle too, we have $\Delta_{i} = \Delta_{i}^{-}+\lidle$ and if it chooses to transmit, its status update will be successfully received and its age at all other nodes will reset to $\Delta_{i}=\lsucc$. Since age of node $i$'s status at other nodes at the beginning of the slot ($\Delta_{i}^{-}$) is at least as large as $\lsucc$, $\Delta_{i}^{-}+\lidle > \lsucc$. Hence, node $i$ will prefer to transmit ($\T$). 

\SG{Case $1$ and Case $2$ shows that player's preferences are opposite, hence, the game has no weakly dominant strategy.}
\end{proof}
\label{prop:lcolgrlsucc}
\end{proposition}
\begin{table*}[t]
\small
\centering
\begin{tabular}{|l|c|p{2cm}|p{4.5cm}|l|}
\hline
Case & $\lcol$ & Vector of Ages seen at the beginning of the stage, $\vect{\Delta}^{-}$ & Mixed Strategy Nash Equilibrium, $\vect{\tau}^{*}$ & Pure Strategy Nash Equilibrium\\
\hline\hline
I & $0.1\lsucc$ & $(\lsucc,2\lsucc,3\lsucc)$ & $(\mathbf{2.4877},\mathbf{-1.2782},0.3549)$ & $(\T,\T,\T),(\T,\I,\T),(\I,\T,\T),(\T,\T,\I)$\\
II & $0.1\lsucc$ & $(\lsucc,\lsucc,\lsucc)$ & $(\mathbf{-0.0055},\mathbf{-0.0055},\mathbf{-0.0055})$ &$(\T,\T,\T),(\T,\I,\T),(\I,\T,\T),(\T,\T,\I)$\\
III & $2\lsucc$ & $(\lsucc,2\lsucc,3\lsucc)$ & $(0.6008, 0.3355, \mathbf{-0.9804})$ & $(\T,\T,\T),(\I,\I,\T),(\T,\I,\I),(\I,\T,\I)$\\
IV & $2\lsucc$ & $(2\lsucc,3\lsucc,3\lsucc)$ & $(0.6008, 0.3355, 0.3355)$ &$(\T,\T,\T),(\I,\I,\T),(\T,\I,\I),(\I,\T,\I)$\\
V  & $2\lsucc$ & $(2\lsucc,3\lsucc,4\lsucc)$ & $(0.6672, 0.5012, 0.0049)$ &$(\T,\T,\T),(\I,\I,\T),(\T,\I,\I),(\I,\T,\I)$\\
\hline
\end{tabular}
\caption{\small Mixed Strategy Nash Equilibrium $\vect{\tau}^{*}$ computed using~(\ref{Eq:strategy}) and the pure strategy Nash Equilibrium corresponding to different selections of $\NA = 3$, $\lcol$ and $\vect{\Delta}$ computed by substituting the values of $\lcol$ and $\vect{\Delta}$ in the payoff matrix shown in Figure~\ref{fig:payoff_3player}. Other parameters used in the computation are $\lsucc = 1.01$, $\lidle = 0.01$.} 
\label{tab:examples}
\end{table*}

We consider mixed strategies and derive the Nash equilibrium when the initial ages of the nodes' updates lie within a certain assumed region. As stated in~\cite{nash}, every finite strategic-form game has a mixed strategy Nash equilibrium (MSNE). For a strategic game $G$, a mixed-strategy profile $\phi^{*} = (\phi_{1}^{*},\phi_{2}^{*},\dots,\phi_{\NA}^{*})$ is a Nash equilibrium~\cite{nash}, if $\phi_{i}^{*}$ is the best response of player $i$ to his opponents' mixed strategy $\phi_{-i}^{*}\in \boldsymbol{\Phi}_{-i}$, for all $i\in\mathcal{N}$. We have 
\begin{align*}
u_{i}(\phi_{i}^{*},\phi_{-i}^{*}) \geq u_{i}(\phi_{i},\phi_{-i}^{*}), \forall \phi_{i} \in \boldsymbol{\Phi}_i,
\end{align*}
where 
$\phi^{*} \in \prod_{i=1}^{|\mathcal{N}|}\boldsymbol{\Phi}_{i}$ is the profile of mixed strategy. 
\SG{As stated earlier, the mixed strategy for player $i$ is identified by the probability $\tau_i \in [0,1]$ with which node $i$ attempts transmission in a slot.} Let $\AvgAoIT{}{}^{-} = ({1}/{\NA})\sum\limits_{i=1}^{\NA}\Delta_{i}^{-}$, which is the average of the ages of nodes' updates at the beginning of the slot.
\begin{proposition}
If $\lcol>\lsucc$ and  
\begin{align}
\AvgAoIT{}{}^{-} - \frac{(\NA-1) \Delta_{i}^{-}}{\NA} >\frac{\lsucc-\lidle}{\NA},\ \forall i\in \mathcal{N} 
\label{Eq:prop2Ineq},
\end{align}
The MSNE is given by 
\begin{align*}
\tau_{i}^{*} = \frac{\lsucc - \lidle +(\NA-1)\Delta_{i}^{-} - \NA\AvgAoIT{}{}^{-}}
{\NA\lsucc - (\NA-1)\lcol - \lidle +(\NA-1)\Delta_{i}^{-}- \NA\AvgAoIT{}{}^{-}},\\
\forall i \in \mathcal{N}. 
\end{align*}
The condition~(\ref{Eq:prop2Ineq}) ensures that $\tau_{i}^{*}\in(0,1)$.
\begin{proof}[\textbf{Proof:}]
For node $i$ to randomize, when other nodes play their mixed strategies $\tau^*_{-i}$, both the pure strategies of transmit and idle must be best responses of $i$. Note that the choice of $\T$ and $\I$ are, respectively, equivalent to setting $\tau_i = 1$ and $\tau_i = 0$ in~(\ref{Eq:agepayoff}). We require $u_{i}(1,\tau^*_{-i}) = u_{i}(0,\tau^*_{-i})$.  Using~(\ref{Eq:agepayoff}) we can write
\begin{subequations}
\begin{align}
&u_{i}(1,\tau^*_{-i}) =\nonumber\\
&- \left(1-\prod\limits_{j\neq i}^{\NA}(1-\tau^*_{j})\right) (\Delta_{i}^{-}+\lcol)-\prod\limits_{j\neq i}^{\NA}(1-\tau^*_{j})\lsucc,\label{Eq:exp_transmit}\\
&u_{i}(0,\tau^*_{-i}) =\nonumber\\
&- \left(1-\prod\limits_{j\neq i}^{\NA}(1-\tau^*_{j})-\sum\limits_{j\neq i}^{\NA}\tau^*_{j}\prod\limits_{k\neq j,k\neq i}^{\NA}(1-\tau^*_{k})\right)\nonumber\\
&\qquad(\Delta_{i}^{-} + \lcol)-\prod\limits_{j\neq i}^{\NA}(1-\tau^*_{j})(\Delta_{i}^{-} + \lidle)\nonumber
\end{align}
\begin{align}
&\qquad-\left(\sum\limits_{j\neq i}^{\NA}\tau^*_{j}\prod\limits_{k\neq j,k\neq i}^{\NA}(1-\tau^*_{k})\right)(\Delta_{i}^{-}+\lsucc).\label{Eq:exp_idle}
\end{align}
\end{subequations}
Equation~(\ref{Eq:exp_transmit}) is explained by the fact that since node $i$ chooses to transmit the slot is either a collision slot or a successful transmission slot. It is the former if one or more of the other nodes transmit. It is the latter in case none of the other nodes transmit. Similarly, Equation~(\ref{Eq:exp_idle}) is explained by the fact that since $i$ chooses to stay idle, the slot can be either idle, a collision slot, or a successful transmission slot, as determined by the probabilistic choices made by the other nodes, which are governed by $\tau^*_{-i}$.

Equating $u_{i}(1,\tau^*_{-i})$ and $u_{i}(0,\tau^*_{-i})$, for node $i$, we obtain 
\begin{align}
\sum\limits_{j\neq i}^{\NA}\frac{\tau_{j}}{1-\tau_{j}} = \frac{\lsucc - \lidle - \Delta_{i}^{-}}{\lsucc-\lcol}.
\label{Eq:cond}
\end{align}
On solving the resulting system of $\NA$ equations, we get the mixed equilibrium
\begin{align}
\tau_{i}^{*} = \frac{\lsucc - \lidle +(\NA-1)\Delta_{i}^{-} - \NA\AvgAoIT{}{}^{-}}
{\NA\lsucc - (\NA-1)\lcol - \lidle +(\NA-1)\Delta_{i}^{-}- \NA\AvgAoIT{}{}^{-}},\nonumber\\
\enspace \forall i \in \mathcal{N}.
\label{Eq:strategy}
\end{align}

Next, we must find the conditions that ensure that $\tau_{i}^{*}$ lies in the interval $(0,1)$. We consider the following two cases: 

\textit{Case I:} When $[\lsucc - \lidle +(\NA-1)\Delta_{i}^{-} - \NA\AvgAoIT{}{}^{-}]>0$ and $[\NA\lsucc - (\NA-1)\lcol - \lidle +(\NA-1)\Delta_{i}^{-}- \NA\AvgAoIT{}{}^{-}]>0$, for $\tau_{i}^{*}>0$ we get:
\begin{subequations}
\begin{align}
\AvgAoIT{}{}^{-} - \frac{\NA-1}{\NA}\Delta_{i}^{-}&< \frac{\lsucc-\lidle}{\NA},\label{Eq:condIa}\\
\AvgAoIT{}{}^{-} - \frac{\NA-1}{\NA}\Delta_{i}^{-}&< \frac{\lsucc-\lidle}{\NA} - \frac{(\NA-1)(\lcol-\lsucc)}{\NA}\label{Eq:condIb}. 
\end{align} 
\end{subequations}
For $\tau_{i}^{*}<1$, we get $[\lsucc - \lidle +(\NA-1)\Delta_{i}^{-} - \NA\AvgAoIT{}{}^{-}]<[\NA\lsucc - (\NA-1)\lcol - \lidle +(\NA-1)\Delta_{i}^{-}- \NA\AvgAoIT{}{}^{-}]$, which on simplification gives $\lsucc>\lcol$. Note that in Proposition~\ref{prop:lsuccgrlcol} we showed that when $\lsucc>\lcol$, the one-shot game $G$ has transmit ($\T$) as a weakly dominant strategy. As a result, we can discard this case, since for no selection of $\vect{\Delta}^{-}$, the inequalities in~(\ref{Eq:condIa}) and~(\ref{Eq:condIb}) will hold true.

\textit{Case II:} When $[\lsucc - \lidle +(\NA-1)\Delta_{i}^{-} - \NA\AvgAoIT{}{}^{-}]<0$ and $[\NA\lsucc - (\NA-1)\lcol - \lidle +(\NA-1)\Delta_{i}^{-}- \NA\AvgAoIT{}{}^{-}]<0$, for $\tau_{i}^{*}>0$ we get:
\begin{subequations}
\begin{align}
\AvgAoIT{}{}^{-} - \frac{\NA-1}{\NA}\Delta_{i}^{-}&> \frac{\lsucc-\lidle}{\NA},\label{Eq:condIIa}\\
\AvgAoIT{}{}^{-} - \frac{\NA-1}{\NA}\Delta_{i}^{-}&> \frac{\lsucc-\lidle}{\NA} - \frac{(\NA-1)(\lcol-\lsucc)}{\NA}\label{Eq:condIIb}. 
\end{align} 
\end{subequations}
For $\tau_{i}^{*}<1$, we get $[\lsucc - \lidle +(\NA-1)\Delta_{i}^{-} - \NA\AvgAoIT{}{}^{-}]>[\NA\lsucc - (\NA-1)\lcol - \lidle +(\NA-1)\Delta_{i}^{-}- \NA\AvgAoIT{}{}^{-}]$, which on simplification gives $\lsucc<\lcol$. Equations~(\ref{Eq:condIIa}),~(\ref{Eq:condIIb}) give us the condition $\AvgAoIT{}{}^{-} - \frac{\NA-1}{\NA}\Delta_{i}^{-} >\max\{\frac{\lsucc-\lidle}{\NA},\frac{\lsucc-\lidle}{\NA} - \frac{(\NA-1)(\lcol-\lsucc)}{\NA}\}$, which implies $\AvgAoIT{}{}^{-} - \frac{\NA-1}{\NA}\Delta_{i}^{-} >\frac{\lsucc-\lidle}{\NA}$.

\end{proof}
\label{prop:msne}
\end{proposition}
\begin{corollary}
The equilibrium strategy $\tau_{i}^{*}$ of node $i$ is a monotonically decreasing function of  $\Delta_{i}^{-}$ and a monotonically increasing function of $\Delta_{j}^{-}$, where, $j\neq i$. 
\begin{proof}
Consider the following derivatives 
\begin{subequations}
\small
\begin{align*}
\frac{\partial\tau_{i}^{*}}{\partial{\Delta_{i}^{-}}} = \frac{(\NA-1)(\NA-2)(\lsucc-\lcol)}{(\NA\lsucc-(\NA-1)\lcol -\lidle + (\NA-1)\Delta_{i}^{-}-\NA\AvgAoIT{}{}^{-})^2}, 
\end{align*}
\begin{align*}
\frac{\partial\tau_{i}^{*}}{\partial{\Delta_{j}^{-}}} = \frac{(\NA-1)(\lcol-\lsucc)}{(\NA\lsucc-(\NA-1)\lcol -\lidle + (\NA-1)\Delta_{i}^{-}-\NA\AvgAoIT{}{}^{-})^2}. 
\end{align*}
\normalsize
\end{subequations}
Clearly, when $\lcol>\lsucc$, $\partial{\tau_{i}}/\partial{\Delta_{i}^{-}}$ is negative and $\partial{\tau_{i}}/\partial{\Delta_{j}^{-}}$ is positive.
\end{proof}
\label{cor:1}
\end{corollary}

Corollary~\ref{cor:1} implies that as a node's age at the beginning of the slot increases, it becomes conservative and access the medium with smaller probability, while the other nodes in the network become aggressive and access the medium with higher probability. As we later show empirically, this reduces the probability of successful transmission for the node with large age while increasing it for the other nodes. As a result, the node with large age sees fewer successful transmissions and its age keeps increasing.


\section{Empirical Evaluation}
\label{sec:results}
In this section, we empirically demonstrate Proposition~\ref{prop:lsuccgrlcol}, Proposition~\ref{prop:msne} and Corollary~\ref{cor:1} using the $3$-player one-shot game. Table~\ref{tab:examples} shows the scenarios when $\tau_{i}^{*}\notin (0,1)$ and when $\tau_{i}^{*}\in (0,1)$ for different selections of $\lcol$ and $\vect{\Delta}$. Since idle slots are much smaller than both collision slots and successful transmission slots, we set $\lidle = 0.01$ and $\lsucc = 1.01$. 
We compute the mixed strategy Nash equilibrium $\vect{\tau}^{*}$ for different scenarios using~(\ref{Eq:strategy}). 

As shown in Table~\ref{tab:examples}, when $\lcol<\lsucc$ (see scenarios I-II), $\tau_{i}^{*}\notin (0,1)$. Note that this is in agreement with Proposition~\ref{prop:lsuccgrlcol} which states that transmit is a weakly dominant strategy when $\lcol<\lsucc$ and hence node would choose to transmit rather than randomize between pure strategies. When $\lcol>\lsucc$, in scenario III, $\tau_{i}^{*}\notin (0,1)$, whereas, in scenarios IV-V, $\tau_{i}^{*}\in (0,1)$. In scenario III, while $\lcol>\lsucc$, the inequality $\AvgAoIT{}{}^{-} - \frac{(\NA-1) \Delta_{i}^{-}}{\NA} >\frac{\lsucc-\lidle}{\NA},\ \forall i\in \mathcal{N}$, is not satisfied. In line with Proposition~\ref{prop:msne}, since both inequalities are not satisfied, $\tau_{i}^{*}\notin (0,1)$. 
In contrast, in scenarios IV-V, since $\lcol>\lsucc$ and the inequality $\AvgAoIT{}{}^{-} - \frac{(\NA-1) \Delta_{i}^{-}}{\NA} >\frac{\lsucc-\lidle}{\NA}$ holds true $\forall i\in \mathcal{N}$, we get 
$\tau_{i}^{*}\in(0,1),\ \forall i \in \mathcal{N}$. 

\begin{figure}[t]
	\begin{center}
		\includegraphics[width=0.6\columnwidth]{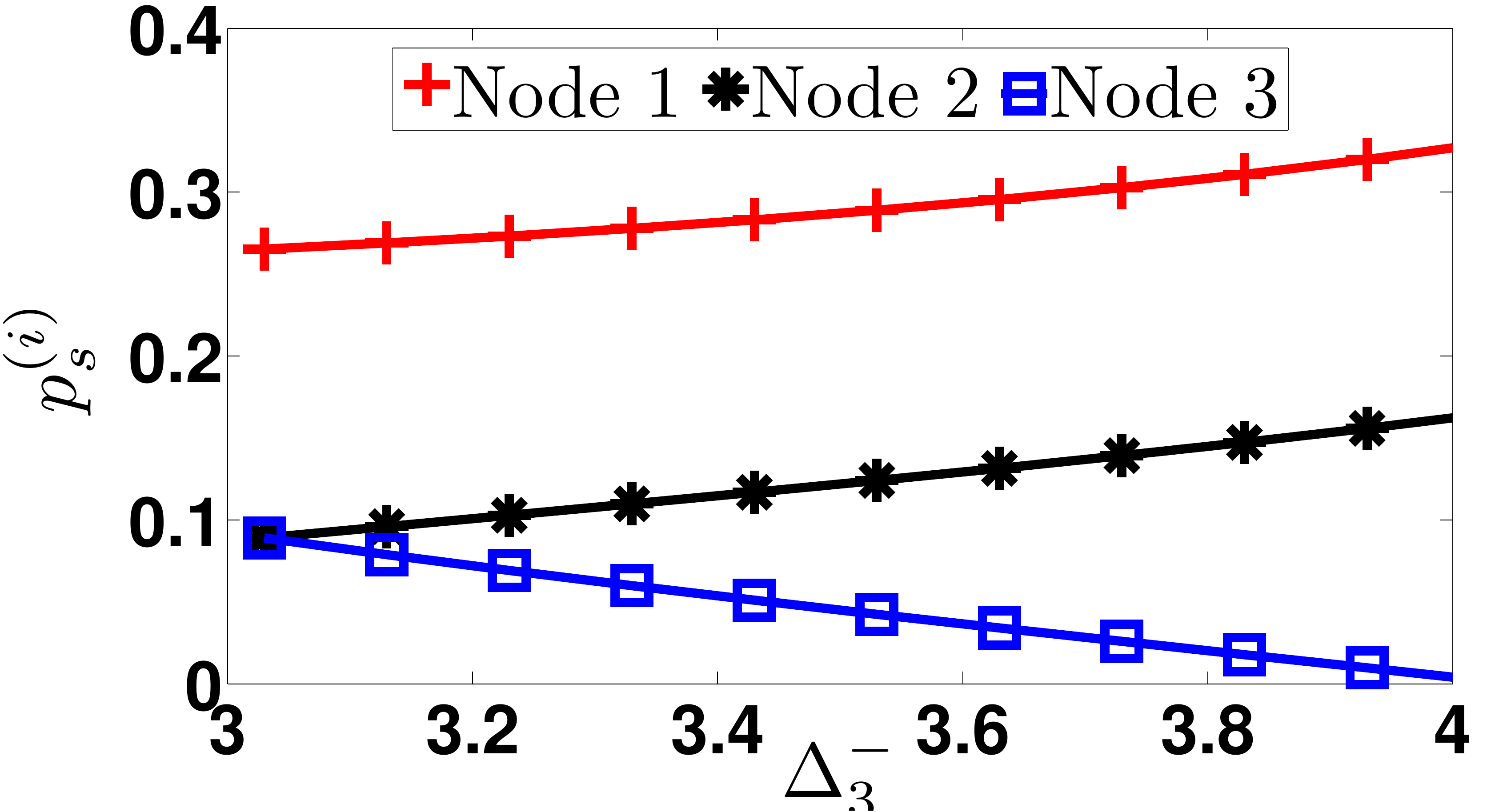}
	\end{center}
	\caption{\small Probability of successful transmission of each node $(\psucci{i})$ for different values of $\Delta_{3}^{-}$, i.e., age of node $3$ at the beginning of the slot, when $\Delta_{1}^{-} = 2\lsucc$ and $\Delta_{2}^{-} = 3\lsucc$.}
	\label{fig:prob}%
	\vspace{-0.33em}
\end{figure}

The equilibrium strategy values in scenarios IV-V are in agreement with Corollary~\ref{cor:1}. 
As shown in Table~\ref{tab:examples}, in scenario IV-V, we can see that as the age of node $3$ at the beginning of the slot increases from $3\lsucc$ to $4\lsucc$ while the age of other nodes is the same, the node becomes conservative and its equilibrium strategy reduces from $0.3355$ to $0.0049$, whereas, node $1$ and $2$ become aggressive and their equilibrium strategy increases from $0.6008$ to $0.6672$ and $0.3355$ to $0.5012$, respectively. This reduces the probability of successful transmission for node $3$, while that for other nodes increases. Figure~\ref{fig:prob} illustrates this for different selections of $\Delta_{3}^{-}$. As a result, node $3$ sees fewer successful tranmissions and its age keeps increasing.

Further, for each scenario we find the pure strategy Nash equilibrium by substituting the values of $\lsucc$, $\lidle$, $\lcol$ and $\vect{\Delta}^{-}$ in the payoff matrix shown in Figure~\ref{fig:payoff_3player}. We see that for all selections of $\lcol$, multiple pure strategy Nash equilibria exist. While $(\T,\T,\T)$ is a common equilibrium strategy for all selections of $\lcol$, we see that when $\lcol<\lsucc$, the equilibrium strategy set comprises of $(\T,\I,\T),(\I,\T,\T),(\T,\T,\I)$ indicating that nodes prefer transmit over idle. This is because when $\lcol<\lsucc$, the age due to a collision ($\Delta_{i} = \Delta_{i}^{-}+\lcol$) is less than that due to a busy slot ($\Delta_{i} = \Delta_{i}^{-}+\lsucc$). Similarly, when $\lcol>\lsucc$, the equilibrium strategy set consists of $(\I,\I,\T),(\T,\I,\I),(\I,\T,\I)$ indicating that nodes prefer idle over transmit, since the age due to a busy slot ($\Delta_{i} = \Delta_{i}^{-}+\lsucc$) would be less than that due to a collision ($\Delta_{i} = \Delta_{i}^{-}+\lcol$).

\section{Conclusion and Future Work}
\label{sec:conclusions}
We formulated a non-cooperative one-shot multiple access game with $N$ nodes as players, where each node shares the spectrum using a CSMA/CA based access mechanism and would like to minimize the age of its updates at the other nodes. We investigated the equilibrium strategies of nodes in a CSMA/CA slot 
for two distinct medium access settings (a) collision are shorter than successful transmissions, and (b) collisions are longer. We showed that when collisions are shorter, transmit is a weakly dominant strategy and when collisions are longer, no weakly dominant strategy exists and under certain conditions on the ages at the beginning of the slot, we derive the mixed strategy Nash equilibrium. 

In the examples illustrated in Table~\ref{tab:examples}, for the values of ages for which the MSNE, as derived analytically, does not lie in the interval $(0,1)$, we see existence of multiple pure Nash equilibrium. This tells us we need to further analyze the game to gain qualitative insight into the equilibria. We leave this analysis for future work.
\begin{spacing}{0.96}
\bibliographystyle{IEEEtran}
\bibliography{references}
\end{spacing}
\end{document}